\documentstyle[epsf,twocolumn,seceq,xspace,bm]{jpsj}

\newcommand{\urs}{URu$_2$Si$_2$\xspace}
\newcommand{\tord}{$T_{\rm o}$\xspace}

\newcommand{\tmag}{$T_{\rm m}$\xspace}
\newcommand{\moment}{$\mu_{\rm o}$\xspace}
\newcommand{\mb}{$\mu_{\rm B}/{\rm U}$\xspace}
\newcommand{\dir}[1]{$[ #1 ]$\xspace}

\title
{
Effects of Uniaxial Stress on Antiferromagnetic Moment\\ in the Heavy Electron Compound URu$_{\bm 2}$Si$_{\bm 2}$
}

\def\runtitle
{
Effects of Uniaxial Stress on Antiferromagnetic Moment in URu$_2$Si$_2$
}

\author
{
Makoto {\sc Yokoyama}, Hiroshi {\sc Amitsuka}, Kenji {\sc Watanabe}$^{1}$, Shuzo {\sc Kawarazaki}$^{1}$,\\ Hideki {\sc Yoshizawa}$^{2}$ and John A. {\sc Mydosh}$^{3}$
}

\def\runauthor
{
Makoto {\sc Yokoyama} {\it et al.}
}

\inst
{
Graduate School of Science, Hokkaido University, Sapporo 060-0810, Japan\\
$^1$Graduate School of Science, Osaka University, Toyonaka 560-0043, Japan\\
$^2$Neutron Scattering Laboratory, Institute for Solid State Physics, University of Tokyo, Tokai 319-1106, Japan\\
$^3$Kamerlingh Onnes Laboratory, Leiden University, P.O.\ Box 9504, 2300 RA Leiden, The Netherlands
}

\recdate
{
September 12, 2001
}

\abst
{
We have performed the elastic neutron scattering experiments under uniaxial stress $\sigma$ along the tetragonal \dir{100}, \dir{110} and \dir{001} directions for \urs . For $\sigma$\,$||$\,\dir{100} and \dir{110}, the antiferromagnetic moment \moment is strongly enhanced from 0.02 \mb ($\sigma=0$) to 0.22 \mb ($\sigma=0.25$ GPa) at 1.5 K. The rate of increase $\partial \mu_{\rm o}/\partial \sigma$ is roughly estimated to be $\sim$ 1.0 $\mu_{\rm B}$/GPa, which is much larger than that for the hydrostatic pressure ($\sim$ 0.25 $\mu_{\rm B}$/GPa). Above 0.25 GPa, \moment shows a tendency to saturate similar to the behavior in the hydrostatic pressure. For $\sigma$\,$||$\,\dir{001}, on the other hand, \moment shows only a slight increase to 0.028 \mb ($\sigma$ = 0.46 GPa) with a rate of $\sim$ 0.02 $\mu_{\rm B}$/GPa. The observed anisotropy suggests that the competition between the hidden order and the antiferromagnetic state in \urs is strongly coupled with the tetragonal four-fold symmetry and the $c/a$ ratio, or both.
}

\kword
{
heavy-electron compound, URu$_{\bf 2}$Si$_{\bf 2}$, uniaxial stress, elastic neutron scattering, antiferromagnetic moment
}

\begin{document}
\sloppy
\maketitle

The nature of the phase transition at $T_{\rm o}$=17.5 K in \urs \cite{rf:Palstra85, rf:Schlabitz86, rf:Maple86} is one of the most fascinating subjects among a variety of phenomena observed in the heavy electron systems. The neutron scattering measurements revealed that the type -I antiferromagnetic structure develops below \tord  \cite{rf:Broholm87,rf:Mason95}. However, the observed staggered moment \moment is extremely small, and unsuited to large bulk anomalies, such as a specific heat jump ($\sim$ 300 mJ/K$^2$ mol) at \tord . This inconsistency drives us to the question whether it is intrinsic to the phase transition or not. To solve the problem, various ideas have been proposed so far, which are based on the magnetic dipolar order with highly reduced $g$ values\cite{rf:Sikkema96,rf:Okuno98,rf:Nieuwenhuys87,rf:Yamagami2000} or some hidden order due to a non-dipolar degree of freedom\cite{rf:Santini94,rf:Ami93,rf:Ohkawa99,rf:Tsuruta2000,rf:Barzykin95,rf:Kasuya97,rf:Ikeda98}.

We recently performed the neutron scattering measurements under hydrostatic pressure\cite{rf:Ami99}, and found that \moment is continuously enhanced from 0.017 \mb ($P=0$) to 0.25 \mb ($P$$\sim$1.0 GPa). In addition, \moment shows a discontinuous jump from 0.23 \mb to 0.4 \mb at $P_{\rm c}\sim$1.5 GPa, indicative of an occurrence of a new antiferromagnetic phase under high pressure. We suggested from these results that the non-magnetic hidden order competes with the antiferromagnetic order in \urs .  Quite recently, the $^{29}$Si-NMR measurements performed under hydrostatic pressures up to 0.83 GPa revealed that the system is spatially divided into two regions below {\tord }: the antiferromagnetic region with a large moment and the non-magnetic one\cite{rf:Matsuda2001}. This also suggests the existence of the hidden order for $P<P_{\rm c}$. Although the origin of the hidden order parameter is still debated, the 5f orbital degree of freedom is regarded as one of the most promising candidates. It is therefore interesting to investigate the relation between the antiferromagnetic moment and the tetragonal crystalline symmetry. We present here for the first time the elastic neutron scattering experiments under uniaxial stress for \urs .

A single-crystalline \urs (the tetragonal ThCr$_2$Si$_2$-type structure) was grown by the Czochralski pulling method in a tri-arc furnace, and vacuum-annealed for a week at 1000$^{\rm o}$C. By means of spark erosion, three plates of dimensions $\sim$ 25 mm$^2$ $\times$ 1 mm were cut from the crystal so that their bases give (001), (100) and (110) planes. The uniaxial stresses $\sigma$ were applied along each \dir{001}, \dir{100} and \dir{110} axis up to 0.46 GPa at room temperature, by putting the samples between piston cylinders (Be-Cu alloy) mounted in a clamp-type pressure cell. The cell was cooled in a $^4$He refrigerator for temperature down to 1.5 K.

The elastic neutron scattering experiments were performed by the triple-axis spectrometer GPTAS (4G) located in the JRR-3M research reactor of Japan Atomic Energy Research Institute. The neutron momentum $k=2.660\ {\rm \AA}^{-1}$ was chosen by using the (002) reflection of pyrolytic graphite (PG) for both monochromating and analyzing the neutron beam. We used the combination of 40'-80'-40'-80' collimators, together with two PG filters to eliminate the higher order reflections. The scans were performed in the ($hk0$), ($h0l)$ and ($hhl$) scattering planes for $\sigma$\,$||$\,\dir{001}, \dir{100} and \dir{110}, respectively. The antiferromagnetic Bragg reflections were obtained by the (100) scans for $\sigma$\,$||$\,\dir{001}, the (100), (102) and (203) scans for $\sigma$\,$||$\,\dir{100}, and  the (111) and (113) scans for $\sigma$\,$||$\,\dir{110}.

Figure 1 shows the uniaxial-stress variations of the elastic longitudinal scans at 1.5 K through the (100) magnetic Bragg peak for $\sigma$\,$||$\,\dir{001} and \dir{100}, and through the (111) peak for $\sigma$\,$||$\,\dir{110}. The instrumental background and the contamination of higher-order nuclear reflections were carefully subtracted by using data at 40 K.  By applying stress along the \dir{100} direction, the (100) peak intensity is strongly enhanced. The (102) and (203) magnetic Bragg peaks also develop rapidly, while no peak is observed at the (001) position. The intensities of (100), (102) and (203) reflections roughly follow the $|Q|$ dependence of the U$^{4+}$ magnetic form factor\cite{rf:Frazer65}. In addition, no other peak was found in scans along the principal axes in the first Brillouin zone. These results indicate that the type-I antiferromagnetic structure with \moment polarized along the $c$ axis is not changed for $\sigma$\,$||$\,\dir{100}. The stress along the \dir{110} direction also increases the intensity of the (111) magnetic Bragg peak. From the same analyses, we confirm that the antiferromagnetic structure is unchanged also for $\sigma$\,$||$\,\dir{110}. For stress along the \dir{001} direction, on the other hand, the intensity of the (100) Bragg peak slightly increases at $\sigma$ = 0.46 GPa, indicating that the antiferromagnetic state strongly depends on the direction of uniaxial stress.
\begin{figure}[tbp]
  \vspace{0cm} 
     \epsfxsize=9cm
	     \centerline{\epsffile{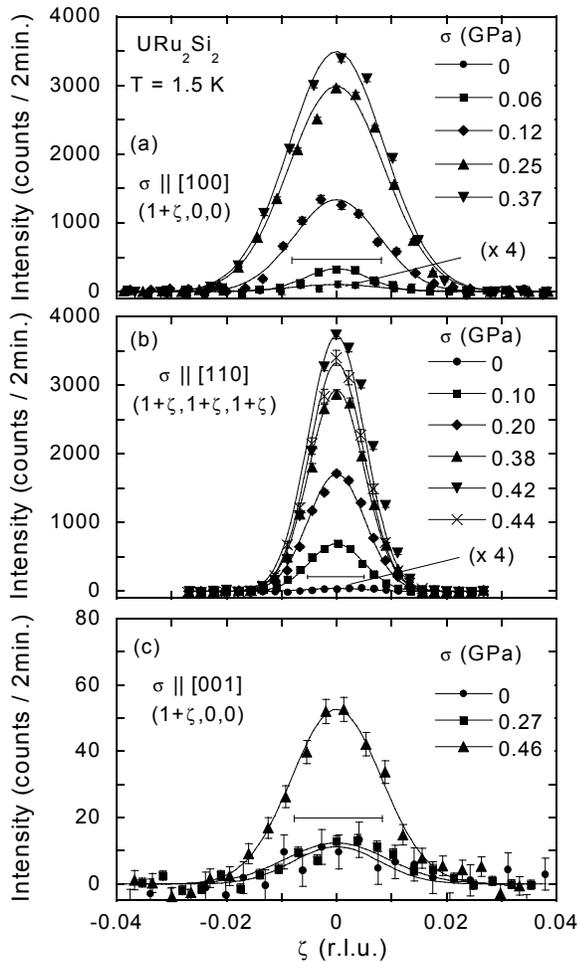}}
  \caption{The uniaxial stress variations of the magnetic Bragg peaks of \urs , observed from the longitudinal scans at (a) the (100) position for $\sigma$\,$||$\,\dir{100}, (b) (111) for $\sigma$\,$||$\,\dir{110}, and (c) (100) for $\sigma$\,$||$\,\dir{001} at 1.5 K. The horizontal bars indicate the widths of the resolution limit estimated from the higher-order nuclear reflections. Note that the data for $\sigma$=0 are 4 times enlarged in (a) and (b).}
\end{figure}

The observed (100) and (111) magnetic Bragg peaks are fitted by the Lorentzian function convoluted with the Gaussian resolution function, to estimate the correlation lengths $\xi$ of \moment . The instrumental resolution are estimated from the widths (FWHM) of higher-order nuclear reflections measured at the corresponding positions without PG filters. At ambient pressure, $\xi$ along the \dir{100}, \dir{001} and \dir{111} directions are estimated to be about 150 ${\rm \AA}$, 260 ${\rm \AA}$ and 330 ${\rm \AA}$, respectively. They increase rapidly by applying stress along the \dir{100} and \dir{110} directions. Above 0.3 GPa, the simple fits give the $\xi$ values of approximately 2.5 times larger than those for $\sigma=0$. For $\sigma$\,$||$\,\dir{001}, on the other hand, $\xi$ for the \dir{100} direction slightly increases to $\sim$ 230 ${\rm \AA}$ at 0.46 GPa, indicating that the increase of $\xi$ is accompanied by the enhancement of \moment .

Displayed in Fig.\ 2 is the uniaxial-stress dependence of \moment at 1.5 K. The magnitudes of \moment are obtained from the integrated intensities of the magnetic Bragg peaks at (100) for $\sigma$\,${||}$\,\dir{001} and \dir{100}, and at (111) for $\sigma$\,${||}$\,\dir{110}, which are normalized by the intensities of the weak nuclear (110) reflection for $\sigma$\,${||}$\,\dir{001} and \dir{110}, and (101) for $\sigma$\,${||}$\,\dir{100}. \moment at $\sigma=0$ is estimated to be 0.020(4) \mb , which roughly corresponds with the values of previous investigations\cite{rf:Broholm87,rf:Mason95,rf:Ami99}. As stress is applied along the \dir{100} direction up to 0.25 GPa, \moment is strongly enhanced to 0.22(2) \mb , and then shows a tendency to saturate above 0.25 GPa. The $\mu_{\rm o}(\sigma$) curve for $\sigma\,||$\,\dir{100} is quite similar to that for the hydrostatic pressure\cite{rf:Ami99}. However, the estimated rate of increase, $\partial \mu_{\rm o}/\partial \sigma$ $\sim$ 1.0 $\mu_{\rm B}/{\rm GPa}$, is much larger than that for hydrostatic pressure, $\partial \mu_{\rm o}/\partial \sigma$ $\sim$ 0.25 $\mu_{\rm B}/{\rm GPa}$. Interestingly, \moment also develops as uniaxial stress is applied along the \dir{110} direction, tracing the curve for $\sigma$\,$||$\,\dir{100} within the experimental accuracy. For $\sigma$\,$||$\,\dir{001}, on the other hand, \moment slightly increases to 0.028(3) \mb at 0.46 GPa, with a small rate $\partial \mu_{\rm o}/\partial \sigma$ $\sim$ 0.02 $\mu_{\rm B}/{\rm GPa}$. \begin{figure}[tbp]
	  \vspace{0cm}
     \epsfxsize=8.5cm
	     \centerline{\epsffile{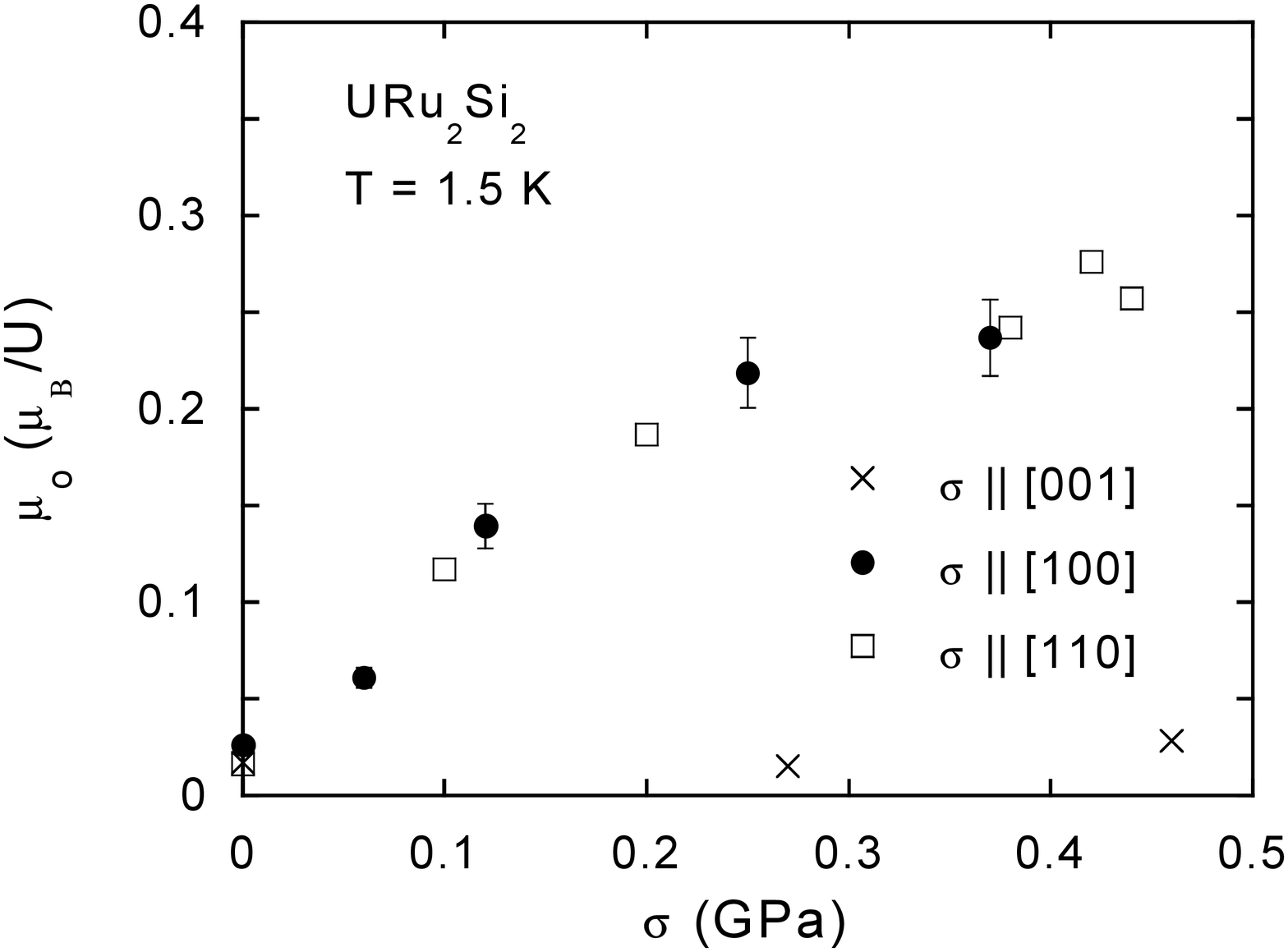}}
  \caption{Uniaxial stress dependence of the staggered magnetic moment \moment at 1.5 K.}
\end{figure}

In fig.\ 3, we plot the normalized Bragg-peak intensity $I/I({\rm 1.5\ K})$ for $\sigma$\,$||$\,\dir{100} and \dir{110} as a function of temperature normalized by the onset of $I(T)$ ($\equiv\,$\tmag ). The onset of $I(T)$ is not sharp. $I(T)$ starts increasing at the temperature $T_{\rm m}^+$ and exhibits a $T$-linear dependence below $T_{\rm m}^-$. The width of this ``tail" of $I(T)$, $\delta T_{\rm m}=T_{\rm m}^+-T_{\rm m}^-$, is estimated to be 2-3 K. We here define \tmag as the midpoint of $T_{\rm m}^+$ and $T_{\rm m}^-$. Although the experimental errors are somewhat large, we can estimate the variations of \tmag with $\sigma$ to be within $\sim\pm1.5\ {\rm K}$ from $T_{\rm m}(\sigma=0)\ \sim$ 17.7 K. This contrasts sharply with the large $\sigma$ variations of \moment . The observed weak variations of \tmag roughly agree with the $\sigma$ variations of \tord ($dT_{\rm o}/d\sigma=1.26\ {\rm K/GPa}$) estimated by the electrical resistivity measurements for $\sigma$\,$||$\,\dir{100}\cite{rf:Bakker92}. For weak stress range below $\sigma \le 0.12\ {\rm GPa}$, $I(T)$ for both the $\sigma$ directions exhibit an unusually slow saturation with decreasing temperature. For further compression, $I(T)$ develops sharper and shows a tendency of more rapid saturation below $T/T_{\rm m}\le 0.4$, pronounced in a rounding of the $I(T)$ curve.
\begin{figure}[tbp]
     \epsfxsize=7.5cm
	     \centerline{\epsffile{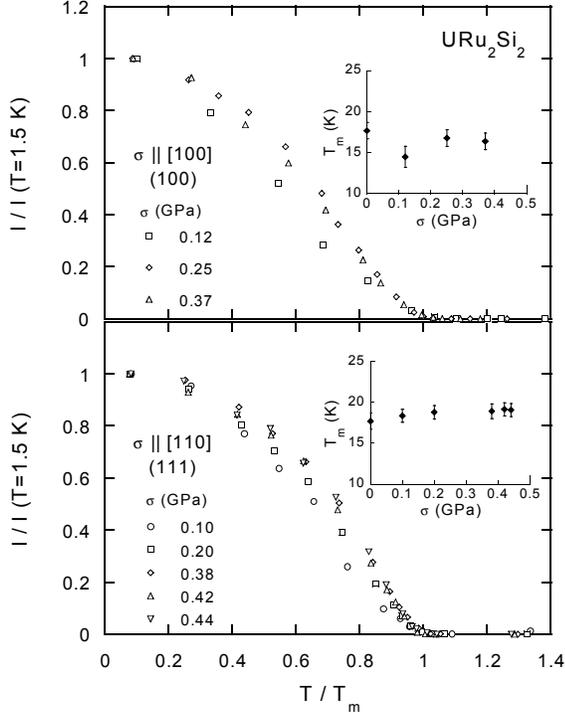}}
  \caption{Temperature dependence of the normalized Bragg peak intensities $I/I(1.5\ {\rm K})$ for $\sigma$\,${||}$\,\dir{100} (top) and $\sigma$\,${||}$\,\dir{110} (bottom). The temperature scales are normalized by the onset temperature of $I(T)$($\equiv$\,\tmag ). The insets show the $\sigma$ dependence of \tmag  .}
\end{figure}

The clear anisotropy observed in the $\sigma$ dependence of \moment suggests that the hidden order is strongly coupled with the tetragonal four-fold symmetry, because it is broken by $\sigma$\,$||$\,\dir{100} and \dir{110}, but preserved for $\sigma$\,$||$\,\dir{001}. Among the existing hidden-order scenarios, the models based on the antiferro-quadrupolar order\cite{rf:Santini94,rf:Ami93,rf:Ohkawa99,rf:Tsuruta2000}, valence transition\cite{rf:Barzykin95}, uranium-pair distortions\cite{rf:Kasuya97}, and the d-type spin density wave\cite{rf:Ikeda98} have been argued in close connection with the tetragonal symmetry. For example, the quadrupolar moments that can order in the tetragonal symmetry are $O_2^2=J_x^2-J_y^2$ and $P_{xy}=\frac{1}{2}(J_xJ_y+J_yJ_x)$, which are expected to be sensitively affected by the uniform distortions of the tetragonal basal plane. All of the above models involve the magnetic instability that the dipolar ($J_z$) order may be caused by destroying the major order, and thus may explain the present experimental results.

Another remark should be made concerning the Poisson's effects. The uniaxial stresses applied along the \dir{100} and \dir{110} directions not only break the tetragonal symmetry but also make the lattice constant $c$ longer. The isotropic feature observed for stresses in the basal plane may suggest that the $c/a$ ratio plays the more important role regarding the competition between the two types of order. In fact, it has been reported that in U(Ru,Rh)$_2$Si$_2$ a large-moment antiferromagnetic order replaces the hidden order by increasing the Rh concentration, coinciding with an increase in $c/a$ ratio\cite{rf:Dalichaouch90,rf:Miyako91,rf:Burlet92}.

In conclusion, we have found strong anisotropy in the uniaxial-stress dependence of the antiferromagnetic state in URu$_2$Si$_2$: the magnetic Bragg-peak intensities increase sharply for stresses along the tetragonal \dir{100} and \dir{110} directions, while the weak variation is observed for stress along the \dir{001} direction. This result implies that the tetragonal four-fold symmetry and the $c/a$ ratio play important role regarding the competition between the hidden order and the antiferromagnetic state in \urs . In contrast to the variations of the scattering amplitude, the transition temperature is nearly independent of stress in any directions.

One of us (M.Y.) is supported by the Research Fellowship of the Japan Society for the Promotion of Science for Young Scientists.

\end{document}